\documentclass{ametsocV6.1}
\usepackage{graphicx}
\usepackage{subfigure}
\usepackage{threeparttable}
\usepackage{siunitx}
\usepackage{url}

\usepackage{enumitem}
\usepackage{lineno}
\nolinenumbers
\renewcommand{\linenumbers}{}
\renewcommand{\internallinenumbers}{}

\title{The Design and Performance of Meteorological Sensors for WindBorne Global Sounding Balloons}

\authors{Jake Spisak,\aff{a}\correspondingauthor{Jake Spisak, jake@windbornesystems.com} 
Christopher P. Riedel,\aff{a} 
Andrey Sushko,\aff{a}
Michal Adamkiewicz,\aff{a}
Joan Creus-Costa,\aff{a} 
John Dean,\aff{a} 
Jacob Radford,\aff{a} 
F. Martin Ralph,\aff{b}
Larissa Reames,\aff{a} 
Anna M. Wilson,\aff{b}
Subin Yoon,\aff{b} 
Vijay Tallapragada, \aff{c}
Todd Hutchinson,\aff{a} 
}

\affiliation{\aff{a}{WindBorne Systems, Palo Alto, CA 94303, USA}\\
\aff{b}{Center for Western Weather and Water Extremes, Scripps Institution of Oceanography, University of California San Diego, La Jolla, CA, USA}\\ 
\aff{c}{NOAA/NCEP Environmental Modeling Center, College Park, Maryland}
}

\abstract{WindBorne Systems has developed a constellation of long-duration atmospheric balloons to collect meteorological data across the globe, filling gaps in current in-situ data collection methods. Each Global Sounding Balloon (GSB) is capable of flying for weeks or months and performing dozens of soundings while measuring pressure, temperature, humidity, and GNSS-derived position, altitude, and wind velocity. This data is transmitted to ground via satellite, processed, and made available within minutes of being collected. The current meteorological sensor package has remained largely unchanged since mid-2024 and has flown on thousands of GSBs totaling over one million hours of flight time. Here we present the design and performance of this sensor package. The custom readout architecture and housing allow for data collection across nearly all in-flight conditions while minimizing sources of bias and noise. Uncertainty is characterized via sounding reproducibility studies and in-house calibration of pressure, humidity, and temperature sensors. The calibration and data processing procedures have been optimized and validated by comparison with external datasets. We present external validation in the form of 1) side-by-side radiosonde launches performed in collaboration with the Center for Western Weather and Water Extremes at the Scripps Institution of Oceanography, which show agreement within expected uncertainty limits, and 2) intercomparison studies with European Centre for Medium-Range Weather Forecasts Reanalysis v5, which show an aggregate root mean square difference of: Geopotential height -- \qty{14}{m}; Pressure -- \qty{0.36}{hPa}; Temperature -- \qty{0.91}{K}; Wind speed \textit{u} -- \qty{2.45}{m/s}; Wind speed \textit{v} -- \qty{2.50}{m/s}; Relative humidity -- 13\%.}

\begin{document}

\maketitle

%
%
%
%
%

%

\section{Introduction}

\begin{figure}
 \noindent\includegraphics[width=0.49\textwidth]{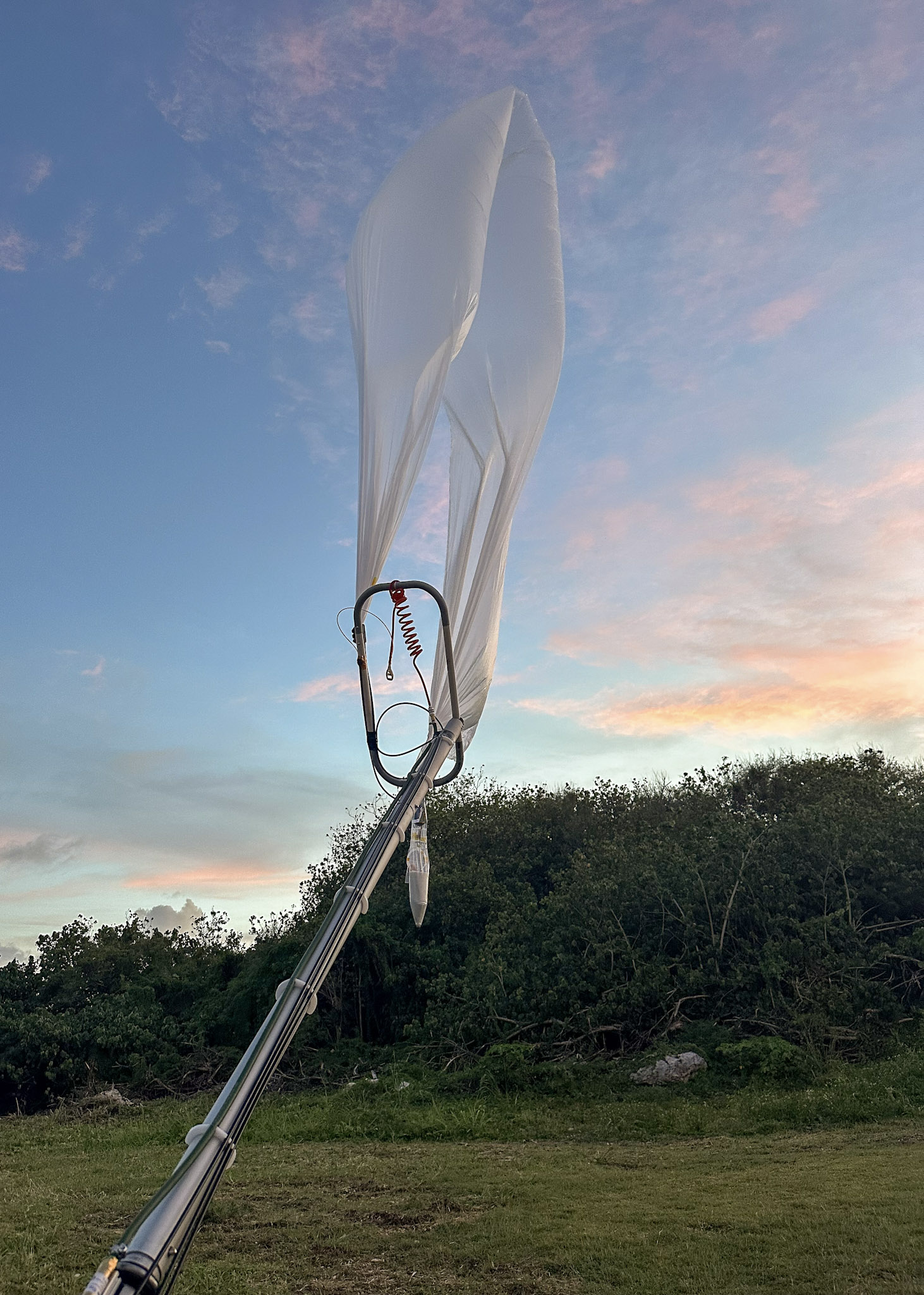}
 \noindent\includegraphics[width=0.49\textwidth]{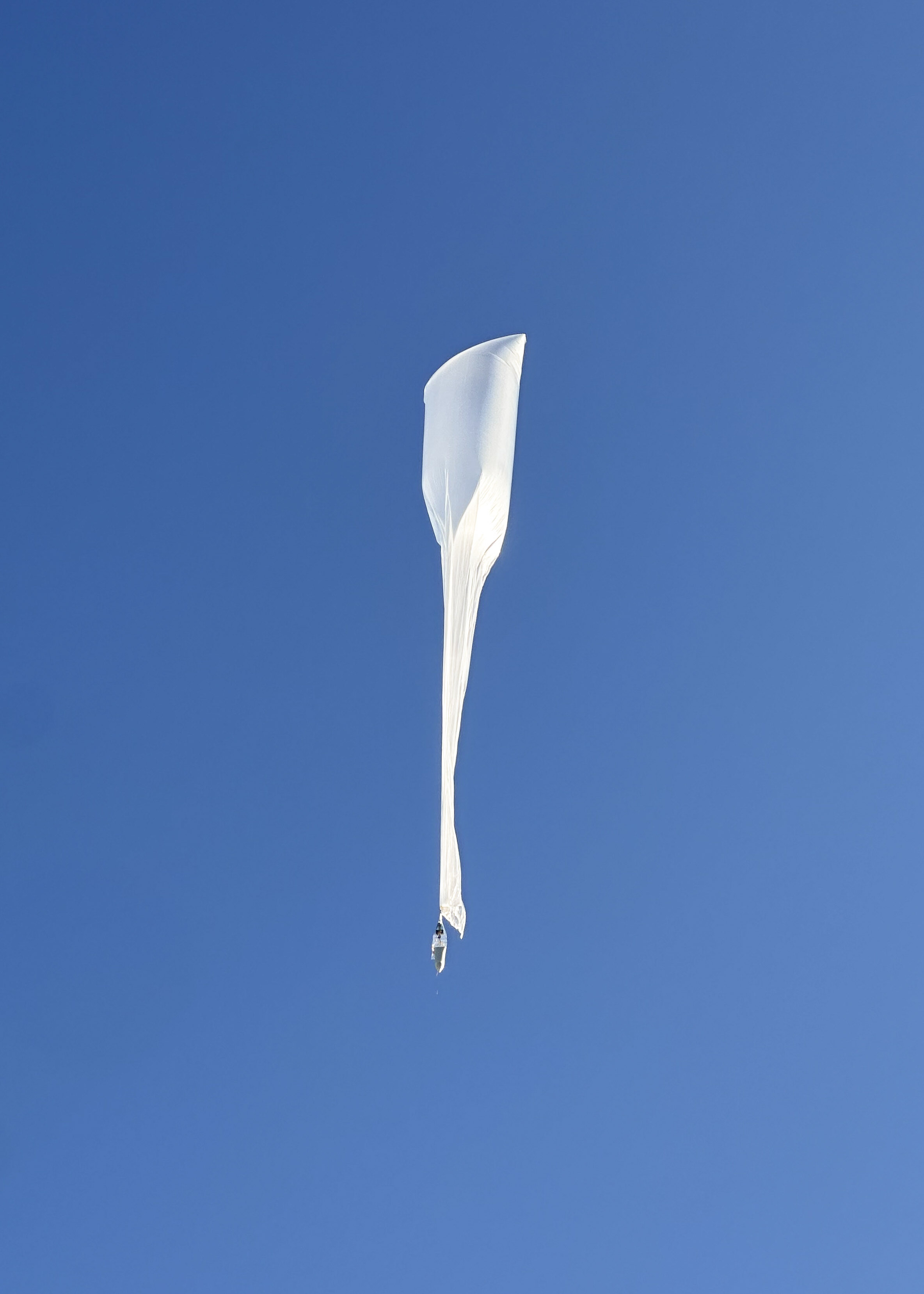}\\
 \caption{\textit{Left:} A GSB on a launch rig, just prior to launch. \textit{Right:} A GSB after launch.}\label{fig:balloon}
\end{figure}

WindBorne Systems has developed the Global Sounding Balloon (GSB), a long-duration controllable balloon-based observation platform optimized for repeated vertical profiling of the atmosphere (Fig. \ref{fig:balloon}). The GSB provides atmospheric observations of pressure, temperature, humidity, and wind speed and direction along its flight path. A single GSB flight is capable of remaining aloft for weeks and collecting dozens of vertical sounding profiles while traveling thousands of kilometers from the launch location. The resulting data most closely resemble radiosonde or dropsonde profiles, but with the hardware costs amortized over dozens of soundings and without the need for a separate delivery vehicle to reach any point on Earth. With near real-time bidirectional satellite communications and fully flexible altitude control, a GSB can be actively navigated by varying altitude to follow horizontal wind trajectories that lead to locations of interest, allowing for targeted observations despite the balloons lacking active propulsion.
\begin{figure}
    \centering
    \includegraphics[width=1\linewidth]{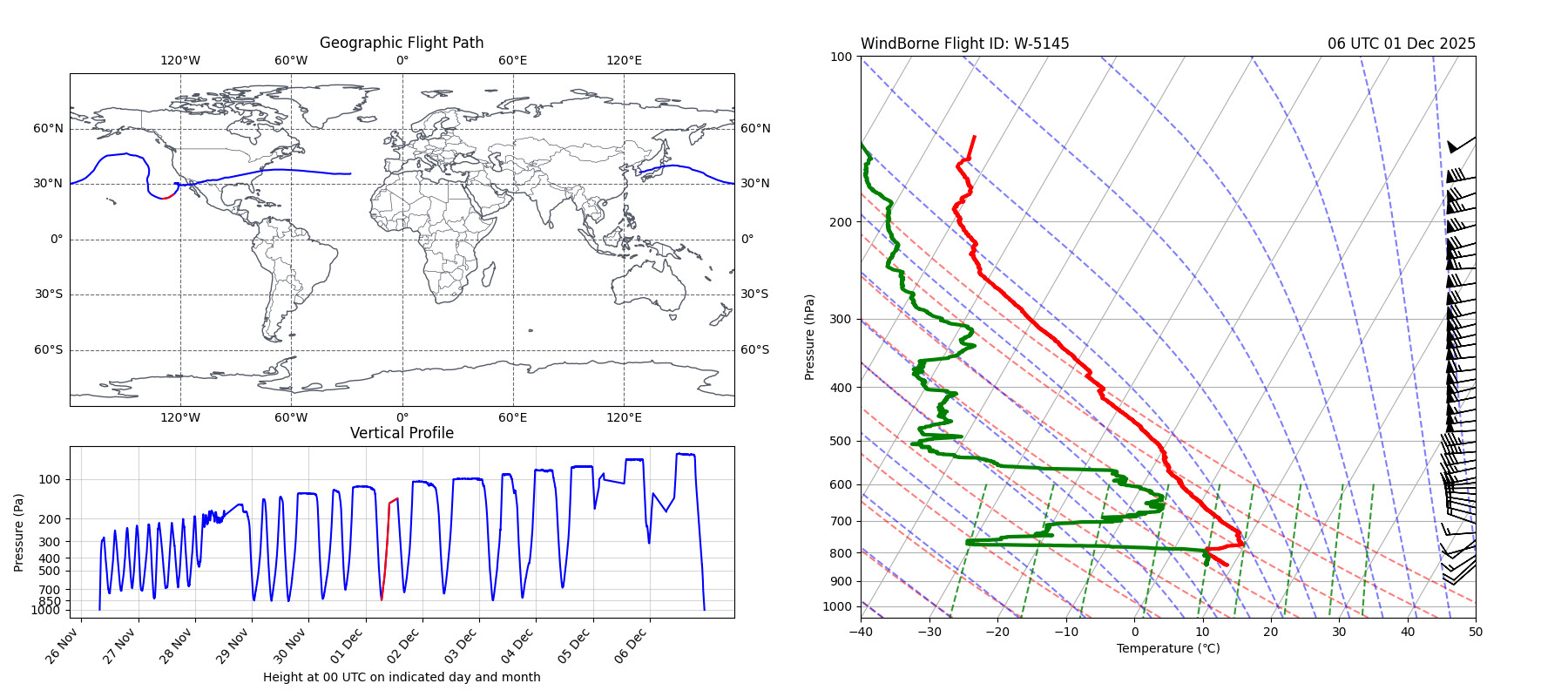}
    \caption{Flight path (upper left) and vertical profile as a function of pressure (lower left) of a GSB that was launched from Uiseong, South Korea on 26 November 2025 and landed on 6 December 2025. Observations of temperature, dew point temperature and wind speed and direction collected by the GSB during an ascent beginning 06 UTC 1 December 2025 are shown on a skew-T, log-P diagram in the right panel.}
    \label{fig:GSB_flight}
\end{figure}

These capabilities are enabled by a radically simple but highly effective approach to altitude control. GSBs are optimized to be a maximally lightweight sensor delivery platform, and operate primarily in a zero-pressure regime below the altitude ceiling. A net positive or negative lift will cause the balloon to rise or fall indefinitely until corrected. Free lift is controlled directly by dropping ballast to reduce weight or venting lift gas to reduce displacement. This operating mode naturally lends itself to repeated profiling as the balloon can be commanded to reduce free lift, initiating a descent from the stratosphere to just above the surface, at which point the free lift is increased, prompting an ascent back to the flight ceiling.  The system can then descend again or trim for neutral lift and remain at relatively constant altitude to travel to another location for the next set of sounding profiles. Crucially, while this approach is intrinsically limited in endurance, it can be accomplished with very cheap and lightweight systems, and scaled down arbitrarily with no loss of performance. Furthermore, adjustments to free lift can be made not only to perform transits through the atmosphere but also to compensate for any perturbations such as ice accumulation, rain, temperature fluctuations, and envelope leakage, leading to performance which degrades gracefully when faced with adverse environmental conditions or envelope defects.

WindBorne currently manufactures and flies two types of GSBs: one optimized for collecting mostly tropospheric vertical profiles of observations (referred to as GSB-60) and one optimized for collecting mostly stratospheric observations (referred to as GSB-80). A typical GSB-60 flight collects profiles that range from between 500 and \qty{2500}{meters} above ground level to the balloon ceiling which increases from \qty{12}{km} to \qty{18}{km} (\qty{\sim200}{hPa} to \qty{\sim75}{hPa}) above mean sea level as the flight lifetime progresses.  A GSB-60 launched from Palo Alto, CA on 11 November 2025 (Fig. \ref{fig:GSB_flight}) traveled across the United States and Atlantic Ocean. As the GSB flew, it was autonomously directed to ascend and descend yielding soundings of atmospheric data. GSB-80 flights (not shown) have a ceiling of \qty{24}{km} (\qty{\sim 30}{hPa}) and are optimized for high-altitude flight rather than frequent vertical profiling and generally remain between the tropopause and \qty{24}{km} above sea level after first ascent.

When GSBs encounter conditions adverse for flight, such as precipitation and icing, they are most often automatically directed to ascend above the adverse layers by releasing ballast. In most instances, GSBs continue flying after encountering adverse conditions but must vent lift gas to compensate for reduced ballast weight. This reduces the potential lifetime of the GSB. To achieve longer duration flights, GSBs are typically directed automatically to avoid areas of expected adverse conditions as predicted by weather prediction models. During much of the study period presented here, GFS forecasts were used to predict areas of adverse conditions. More recently, WindBorne's own weather prediction system, \textit{WeatherMesh} as described in \cite{Duetal2025} has been used to predict adverse conditions.

\begin{figure}
 \noindent\includegraphics[width=0.9\textwidth]{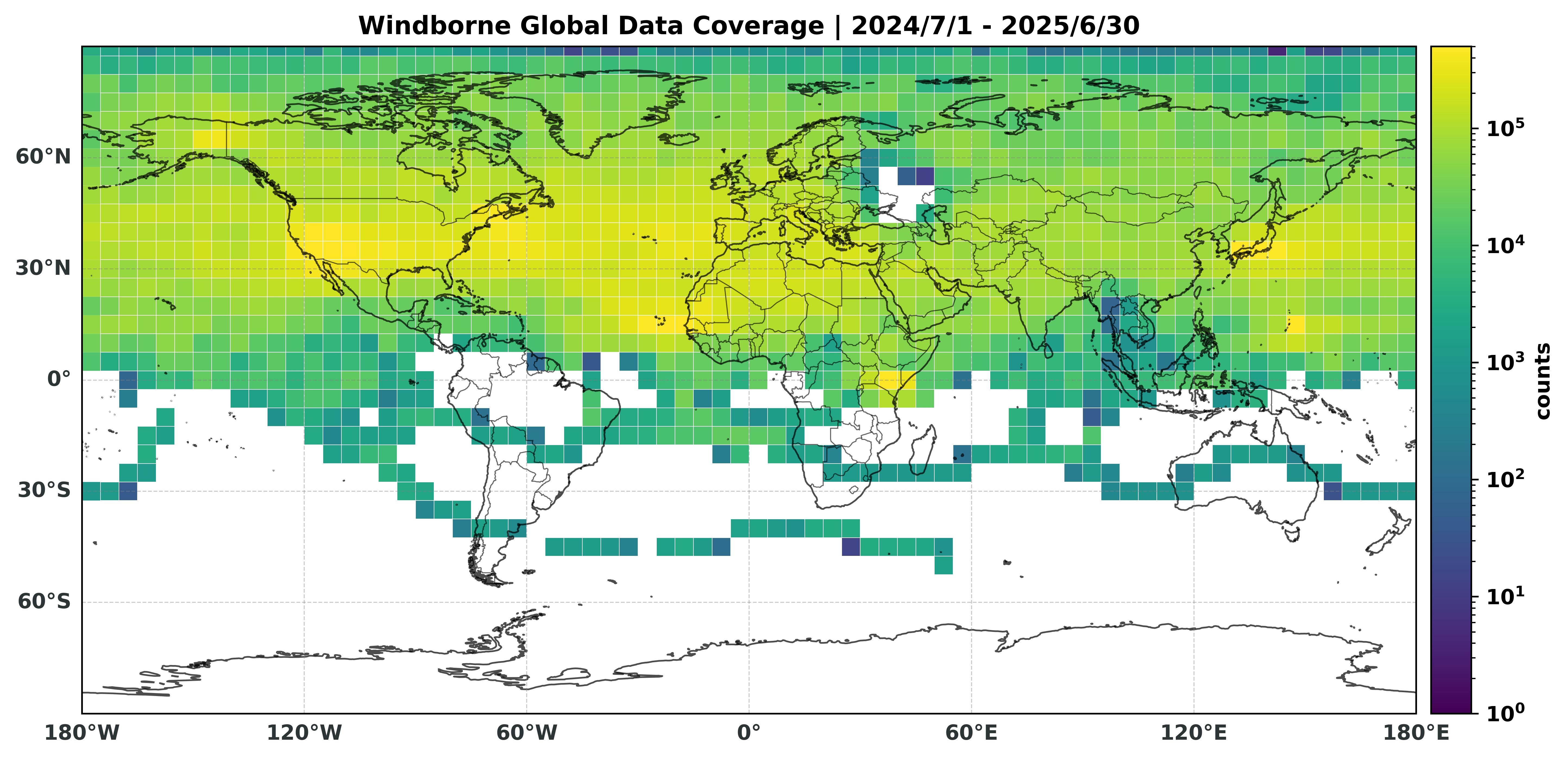}\\
 \caption{Global coverage of WindBorne balloon observations from 1 July 2024 through 30 June 2025. Observation counts are computed by binning 10 second data points onto a 5$^\circ$ $\times$ 5$^\circ$ latitude-longitude grid.}\label{fig:spatialCov_Varcounts}
\end{figure}

In early 2023 WindBorne began operating a constellation of GSBs, referred to as Atlas, by continually launching new flights to replace terminated flights and grow the constellation size.  Between 2019 and December 2025, more than 5500 GSBs have flown and collected atmospheric observations (Fig. \ref{fig:spatialCov_Varcounts}). GSBs have been flown throughout much of the world, with broad coverage over the Northern Hemisphere and reaching remote locations of the Arctic, the northern Pacific and Atlantic oceans and northern Africa. Atlas has grown to an average of 90 GSBs aloft as of December 2025. During November 2025, GSBs flew an average of 3800 vertical km in aggregate per day. Atlas continues to expand and is expected to quadruple in size by the end of 2026.

\section{Atmospheric Sensors}\label{sec:atmospheric_sensors}

\begin{figure}[t]
 \noindent\includegraphics[width=\textwidth]{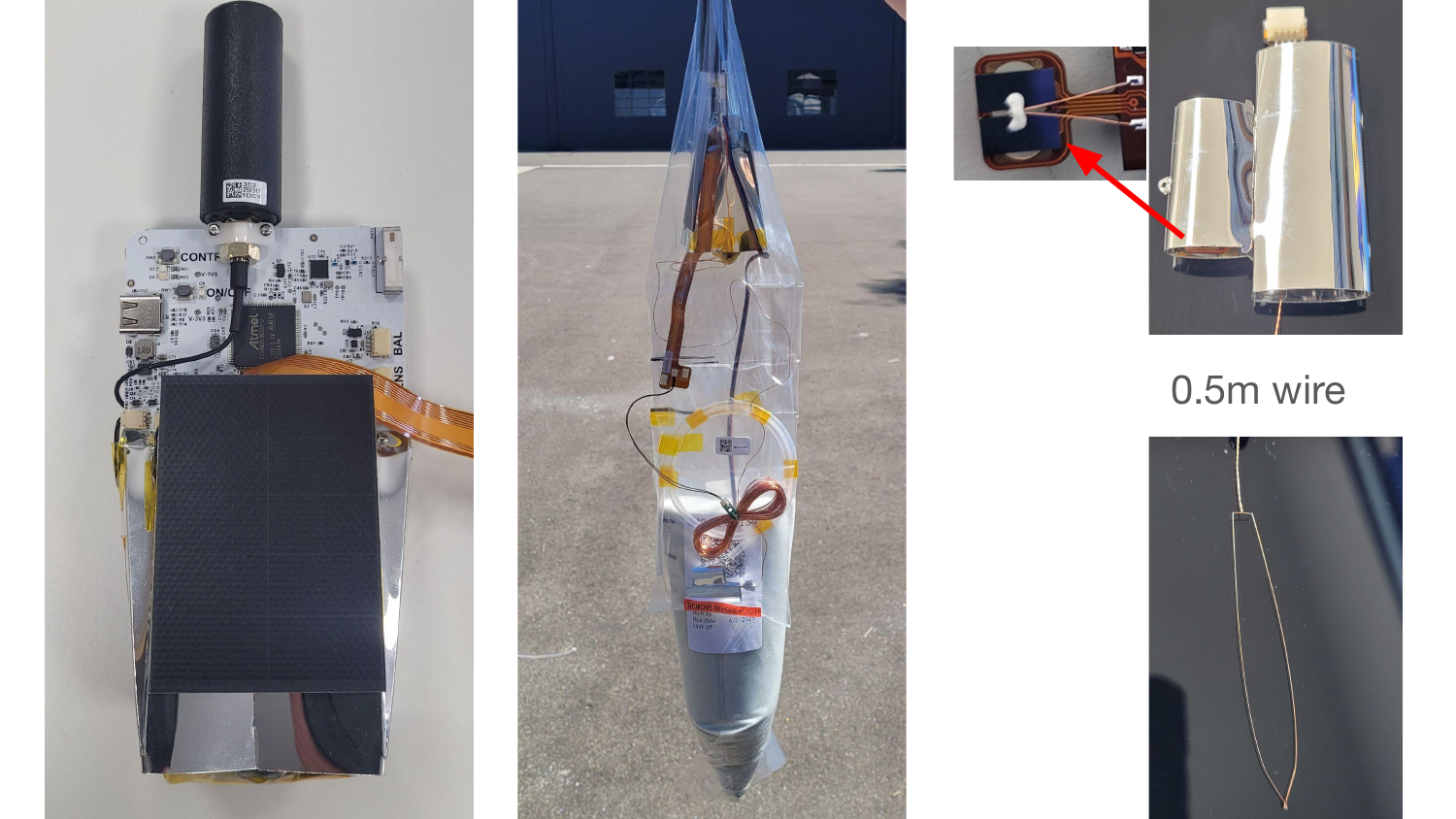}\\
 \caption{\textit{Left:} Main electronics, which include the pressure sensors (not visible) and GNSS module, as well as solar panels, battery, communications, balloon control electronics, and shielding. \textit{Center:} The complete unit attached to a balloon. The main electronics are near the top, and the temperature and humidity sensors are on the board attached to the figure-8 wire loop. After launch, this wire automatically unspools, dropping the sensors \qty{9}{m} below the balloon. \textit{Right:} The humidity sensors, temperature sensor, and their readout electronics are enclosed in an aluminized mylar shield. The two humidity sensors, onto which an additional temperature sensor is epoxied, are located (red arrow) in a shielding configuration that allows for vertical airflow. The atmospheric temperature sensor lies \qty{0.5}{m} below the board on a thin wire.} \label{fig:main_overview} 
\end{figure}

WindBorne GSBs measure temperature, humidity, pressure, altitude, and wind velocity, as well as generate accurate timing and positional data. The sensors are shown in Fig. \ref{fig:main_overview} and performance metrics are in Table \ref{tab:sensor_specs}.

\subsection{Pressure}
The balloon pressure sensor is a Micro-Electro-Mechanical Systems (MEMS) absolute pressure sensor. There are four sensors per balloon located on a custom printed circuit board, which is housed with the main electronics located just underneath the balloon. It has an onboard temperature sensor and provides temperature-compensated pressure readout. To provide meterological-grade accuracy across the necessary temperature range, we perform a calibration procedure that provides a correction to the built-in calibration. The sensors are cycled in a temperature controlled vacuum chamber from \qty{20}{hPa} to \qty{1000}{hPa} and \qty{-60}{\celsius} to \qty{35}{\celsius}, and the pressure readings are compared to an aggregate of three reference pressure transducers, each accurate to \qty{0.1}{hPa} over the calibration range. The offset is then fit as a function of temperature and pressure to generate a calibration correction, typically \qty{<1}{hPa}, for each sensor individually. Post-calibration uncertainty of about \qty{\pm 0.2}{hPa} is driven by a combination of reference uncertainty and hysteresis. This, and other calibration uncertainty numbers, are quoted at the $2 \sigma$ level (two standard deviations). The final pressure measurement provided is a combination of readings from each of the four sensors, with an on-board algorithm that discards outlier readings.

\subsection{Temperature}
The temperature sensor is a negative temperature coefficient thermistor. The thermistor element is composed of metal oxides sintered into a ceramic body, enclosed in a small glass bead approximately \qty{0.7}{mm} in diameter with a time constant of \qty{1}{s} in still air. This time constant is short enough relative to the \qty{10}{s} readout cadence that we do not need to correct for it. The resistance is read out via a custom circuit board suspended by a \qty{9}{m} wire, which prevents readings from being affected by the local environment of the balloon. The wire is coiled prior to launch, and drops after the balloon ascends $\approx 20$--\qty{30}{m}. The thermistor dangles a further \qty{0.5}{m} below the board on a thin wire to similarly prevent the board from affecting the reading. The thermistor is covered in a hydrophobic coating to provide protection from condensation and evaporative cooling. The dominant source of environmental noise is solar radiation, which must be subtracted out as described in Sec. \ref{sec:data_processing}.

The resistance measurement is converted to temperature using a calibration curve provided by the thermistor manufacturer. We additionally perform a custom calibration procedure, after integration into the board, on a sample of the thermistors from each batch. This procedure involves inserting the thermistors into holes within an aluminum block, heat sunk via thermal paste, thermally isolating the block in a temperature chamber, and stepping through the temperature range \qty{-80}{\celsius} to \qty{40}{\celsius} in \qty{10}{\celsius} increments. The results are compared to two reference readings from individually-calibrated platinum Resistance Temperature Detectors, mounted similarly into the block and each accurate to \qty{0.023}{K} over the full calibration range. As a result of this procedure we apply a \qty{<0.3}{\celsius} correction to the calibration curve, with the largest correction applied at lower temperatures. This correction is not sensor-specific, as we find that the spread between individual thermistors is small, typically less than \qty{0.1}{\celsius}, and the batch to batch variance is negligible. This is consistent with the manufacturer's specifications.

\subsection{Humidity}
The humidity sensor is manufactured by E+E Elektronik \footnote{\url{https://www.epluse.com/}} and combines a thin-film capacative relative humidity sensor and a resistive heater on a silicon substrate. There are two humidity sensors per balloon, located on the same custom circuit board that provides temperature readout. Ensuring atmospheric relative humidity measurements are accurate is arguably the most challenging part of the sensor suite --- sensor readings can be biased by contamination, time lag, condensation or icing, poor airflow, and inaccurate temperature readings. The sensors themselves must be carefully handled to avoid damaging the active area, which is exposed on the board. This drove a number of design decisions in manufacturing, calibration, packaging, flight, and data processing. The data processing concerns are described in Sec. \ref{sec:data_processing}, and the remaining design considerations are outlined below.

The two sensors are mounted side by side on a stem offset from the main portion of the board, mounted over an oval cutout to allow exposure to air. The rim of the cutout is gold-plated to avoid moisture accumulation. In order to minimize noise it is desirable that the sensor temperature not exceed the atmospheric temperature by too large an amount, and so the entire board, including the sensors, is covered by an aluminized mylar shield. The shield allows unimpeded vertical airflow past the sensors, except for a small, angled portion of the shield, located directly above sensors to prevent falling ballast from causing damage. A thermistor, identical to the ambient temperature sensor, is epoxied to the back of the humidity sensors to measure their temperature. The built-in heater is automatically run at high relative humidity, via an on-board control loop, to prevent or burn off excess condensation or ice. 

The capacitance measurement is converted to a relative humidity measurement, with respect to water and at sensor temperature, via a temperature-dependent fit provided by E+E Electronik. The fit constant and slope must be individually determined for each sensor via a calibration procedure. Immediately prior to calibration the sensors are washed with deionized water in an ultrasonic cleaner, then reconditioned and rehydrated according to the manufacturer's instructions.  They are then placed in a temperature and humidity chamber at \qty{12}{\celsius} and \qty{25}{\celsius} and several different humidity setpoints ranging from 15\%\,RH to 80\% RH, and the capacitance readings are compared to humidity values provided by two references, each accurate to 1\%\,RH over the calibration range. Fans in the chambers ensure constant airflow past the sensors, and care is taken to minimize the effect of humidity gradients within the chamber. The residual fit error after calibration is typically $<2\%$\,RH. While the calibration fit provided by E+E extends to \qty{-80}{\celsius}, measurements become increasingly difficult at low temperatures and high altitudes. The 2\%\,RH calibration uncertainty is quoted above \qty{-58}{\celsius}, the lowest temperature at which the time constant has been measured (see Sec. \ref{sec:data_processing} \ref{ssec:tau}. After calibration, the sensors are stored in a clean environment and promptly placed in a protected package, which is only opened immediately prior to launching the GSB. Calibration is the last of several manufacturing steps that take place within an environment intended to minimize contamination or damage to the humidity sensors. 

\subsection{Altitude and Wind Velocity}
Altitude and wind velocity are provided by a Global Navigation Satellite System (multi-GNSS) receiver. This also generates latitude, longitude, and timing information. It is located on the main electronics board, along with custom support electronics including a temperature-compensated crystal oscillator. The altitude measurement is reported relative to the Earth Graviational Model 1996 (EGM96; \cite{lemoine1998development}). The reported wind velocity is the balloon velocity --- the location of the GNSS receiver immediately below the balloon means that there is no need to correct for a 'pendulum effect', the motion of the receiver relative to the balloon envelope. The wind speed in each direction is calculated on-board up to \qty{100}{m/s}, and for values that exceed this threshold, is calculated ground-side in data processing using GNSS location and time. The \qty{100}{m/s} maximum was set due to the trade-off between data range and resolution, and allows the vast majority of wind speeds to be calculated on-board and downlinked at sufficient resolution.

\begin{table}[h]
\centering
\begin{threeparttable}
\caption{Sensor Performance Metrics}
\label{tab:sensor_specs}
\begin{tabular}{ll}
\topline
\textbf{Pressure} & \textbf{MEMS}\\
Operating range: & 20 to \qty{1050}{hPa}, -60 to \qty{40}{\celsius}\tnote{a}\\
Data resolution & \qty{0.03}{hPa}\\
Calibration uncertainty & \qty{0.2}{hPa}\\
Sounding reproducibility & \qty{0.25}{hPa}\tnote{b}\\
\midline
\textbf{Temperature} & \textbf{Thermistor}\\
Operating range & -90 to \qty{40}{\celsius}\\
Thermal time constant & $\tau_{63}=$\qty{1}{s} in still air\\
Data resolution & \qty{0.3}{\celsius} \\
Calibration uncertainty & \qty{0.1}{\celsius}\\
Sounding reproducibility & \qty{0.25}{\celsius}\tnote{b}\\
\midline
\textbf{Humidity} & \textbf{E+E Elektronik}\\
Operating range & -80 to \qty{60}{\celsius}, 0 to 100\%\,RH\\
Response time & $\tau_{63}=$\qty{0.9}{s} at \qty{0}{\celsius}, \qty{20}{s} at \qty{-40}{\celsius}\\
Data resolution & 0.5\%\,RH\\
Calibration uncertainty & 2\%\,RH\\
Sounding reproducibility & 2.5\%\,RH\tnote{b}\\
\midline
\textbf{Altitude} & \textbf{Multi-GNSS}\\
Data resolution & \qty{8}{m} \\
Sounding reproducibility & \qty{3}{m} \tnote{c}\\
\midline
\textbf{Wind Speed} & \textbf{Multi-GNSS}\\
Operating range & $<|100|$\,m/s onboard, $\geq |100|$\,m/s calculated ground-side\\
Data resolution & \qty{0.4}{m/s}\\
Sounding reproducibility & \qty{0.4}{m/s}\tnote{b}\\
\botline
\end{tabular}
\begin{tablenotes}
\footnotesize
\item[a] Main electronics, including pressure sensors, are always heated above \qty{-60}{\celsius} during flight
\item[b] One standard deviation difference between simultaneously launched GSBs, matched at equivalent altitude
\item[c] One standard deviation difference between simultaneously launched GSBs, matched at equivalent pressure
\item[] Humidity uncertainty specifications apply above \qty{-58}{\celsius}
\item[] All calibration uncertainties are quoted at $2 \sigma$
\end{tablenotes}
\end{threeparttable}
\end{table}

\section{Communications and Data Processing}\label{sec:data_processing}
A single data point is generated every \qty{10}{s} on-board the GSB, representing the average sensor reading over the \qty{10}{s} interval. Note that during a typical profile a GSB has a vertical speed of \qty{1}{m/s}, in which case this represents one data point every $\approx10$ vertical meters. These datapoints are bundled together into a transmission that is downlinked to ground via the Iridium satellite network\footnote{https://www.iridium.com/} at a configurable interval, typically set to somewhere between 1 and 7 minutes. The flight data is then processed by WindBorne servers and made available. The end-to-end time between data collection and delivery is normally less than 15 minutes. The final output is the following set of variables: Altitude (with respect to EGM96), pressure, wind speed \textit{u} (east/west, eastward positive), wind speed \textit{v} (north/south, northward positive), temperature, relative humidity, latitude, longitude, pressure validity, temperature validity, humidity validity, GNSS validity (altitude, wind speed \textit{u} and \textit{v}, latitude, longitude). The remainder of this section describes certain key aspects of the data processing pipeline. Post-processed performance metrics are shown in Table \ref{tab:sensor_specs}. 

\subsection{Humidity Processing}\label{ssec:humidity_processing}
Readings from the two sensors are averaged to form a single reading. If one sensor's readings do not pass quality control, then the single valid sensor reading is used. The raw measurement is reported at the sensor temperature, with respect to vapor pressure over liquid water. Our post-processed measurement is reported at ambient temperature $T_{\text{amb}}$, with respect to vapor pressure over liquid water for $T_{\text{amb}} \geq \qty{0}{\celsius}$, over ice for $T_{\text{amb}} < \qty{-23}{\celsius}$, and interpolated for $-23 < T_{\text{amb}} < \qty{0}{\celsius}$ following formula 3.12 in \cite{ecmwfifs}. The vapor pressure formulae are taken from \citealt{hyland1983formulations}.  The applied correction is then
\begin{align}
\text{Corrected RH} &= \text{Original RH} \times \frac{e_{\text{w}}(T_{\text{sensor}})}{e_{\text{w}}(T_{\text{amb}})} \times \frac{e_\text{{w}}(T_{\text{amb}})}{e_{\text{w/i}}(T_{\text{amb}})}\label{eq:corrected_rh}.
\end{align}
where $e_{\text{w}}$ is the saturation vapor pressure over liquid water, $e_\text{{w/i}}$ is the combined water/ice saturation vapor pressure, and $T_{\text{sensor}}$ is the humidity sensor temperature. This correction can become large when the sensor and ambient temperature different significantly, primarily when the sun is up and vertical airflow is low. At a certain threshold the converted humidity value becomes too noisy and must be rejected, as described further in Sec. \ref{sec:data_processing}\ref{ssec:quality_control_checks}.

\subsection{Humidity Sensor Time Lag Correction}\label{ssec:tau}
The E+E Electronik humidity sensor was selected for its fast response time, which is \qty{0.9}{s} at \qty{0}{\celsius} and \qty{20}{s} at \qty{-40}{\celsius}. Furthermore, a typical WindBorne GSB sounding is approximately $5 \times$ slower than that of a traditional radiosonde, which means that when measuring a vertical humidity gradient, the GSB error generated by the time lag is significantly less than radiosonde error. Nevertheless, at low temperatures the time lag has a noticable impact on humidity measurements and must be corrected for. For an exponential humidity sensor response, the change in the measured value is given by 
\begin{equation}
    \frac{d\text{RH}_{\text{measured}}}{dt} = (\text{RH}_\text{{true}} - \text{RH}_\text{{measured}}) / \tau.
\end{equation}
Here $\text{RH}_{\text{true}}$ is the true humidity, $\text{RH}_{\text{measured}}$ is the humidity measured by the sensor, and $\tau$ is the time constant. For a perfectly exponential response, $\text{RH}_{\text{measured}}$ will change by $63\%$ in time $\tau$ given a step change in $\text{RH}_{\text{true}}$. In principle, we know $\tau$ and measure both $\frac{d\text{RH}_{\text{measured}}}{dt}$ and $\text{RH}_{\text{measured}}$, allowing us to calculate $\text{RH}_{\text{true}}$. In practice, there are a number of complications, including sensor noise, data resolution, and uncertainty in $\tau$. 

$\tau$ is a strong function of temperature and has been directly measured, with uncertainty, at five points by E+E Electronik. We interpolate in a log-linear fashion $1\sigma$ below the measured values and conservatively extrapolate below the minimum measured temperature of \qty{-58}{\celsius}. Below this temperature, the time lag correction has additional uncertainty. The value of $\tau$ constructed in this way is designed to err on the side of undercorrection rather than overcorrection.

The impact of data resolution is particularly important when $\text{RH}_{\text{measured}}$ changes sufficiently slowly, as the resulting `stair steps' in $\text{RH}_{\text{measured}}$ lead to large uncertainty in the value of $\frac{d\text{RH}_{\text{measured}}}{dt}$. The choice of smoothing algorithm is crucial to accurately recovering $\text{RH}_{\text{true}}$ in these regimes. To estimate $\text{RH}_{\text{true}}$ we use the procedure proposed in \cite{Miloshevich2004}, hereafter Miloshevich 2004. It was developed for Vaisala radiosondes which employed thin-film capacitive humidity sensors with time constants similar to ours, and was validated against a NOAA reference-quality cryogenic hygrometer. 

The procedure involves reducing the dataset to a single point per stair step, applying a smoothing algorithm, and then resampling to the original time steps. The smoothing algorithm is an iterative procedure that seeks to minimize the third derivative of RH, subject to the constraint that each $\text{RH}_{\text{measured}}$ datapoint not change by more than a specified tolerance. We choose a tolerance value that ensures that the smoothing changes $\text{RH}_{\text{measured}}$ by no more than $\pm 0.15\%$\,RH, which is consistent with Miloshevich 2004 when adjusted for differences in resolution. Miloshevich 2004 notes that uncertainty in smoothing choice can generate errors of up to several percent RH in low-temperature environments with strong humidity gradients. That uncertainty should be significantly reduced for the vast majority of WindBorne data due to the slower ascent rates and higher resolution.

\subsection{Temperature Sensor Radiative Bias Correction}
\label{sec:radiative_bias}
The temperature sensor may be biased from reading the true ambient temperature by additional sources of heat transfer, including conduction along the wires, self heating from the measurement, and radiation. In practice we find that solar radiation provides by far the most important source of error. If not corrected, during the daytime the temperature sensor will read about 1--\qty{2}{\celsius} hotter than ambient in typical conditions. The correction function relies on predicting and subtracting this solar bias. The bias is modeled as a heat transfer balance problem,
\begin{equation}
    h_{\text{convective}} \Delta T = q_{\text{solar}},
\end{equation}
where $h_\text{{convective}}$ is the convective heat transfer coefficient, $q_\text{{solar}}$ is the absorbed solar heat flux, and $\Delta T$ is the solar bias. $h_\text{{convective}}$ is based on the convective heat transfer model in \cite{Whitaker1972} for flow past a sphere, and $q_\text{{solar}}$ is determined by calculating the solar flux at a given location in the atmosphere. In total, the computed bias is a function of air density, ascent rate, temperature, solar elevation, and altitude. The bias correction is zero after sunset, as determined by the solar elevation at balloon altitude. 

\begin{figure}[t]
 \noindent\includegraphics[width=\textwidth]{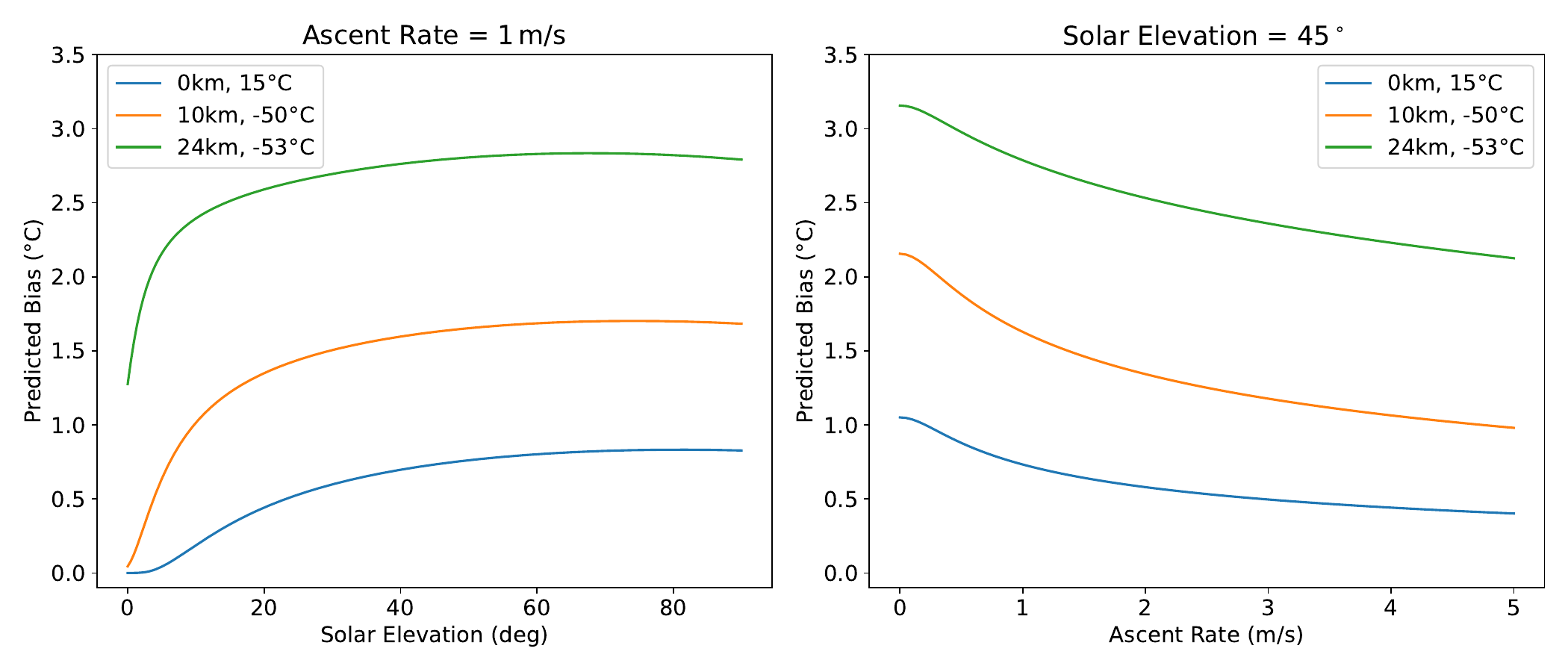}\\
 \caption{The predicted solar bias, which is a function of air density, ascent rate, temperature, solar elevation, and altitude. \textit{Left}: Bias vs solar elevation (at sea level) at several altitudes and standard atmosphere temperatures, at a fixed ascent rate of \qty{1}{m/s}, typical for a profiling GSB. \textit{Right}: Bias vs ascent rate at a fixed solar elevation of $45^\circ$. }\label{fig:solar_bias_vs_params}
\end{figure}

The calculation also requires five empirically-derived coefficients, such as the solar absorptivity of the bead. These were determined by comparing GSB flight data from 400 flights, across all conditions experienced by the balloon constellation, to ERA5 data and performing a fit. The resulting solar bias subtracted is shown in Fig. \ref{fig:solar_bias_vs_params} over a range of parameters. The typical solar bias when profiling at \qty{1}{m/s} is 0.5 -- \qty{1.5}{K} below \qty{10}{km} and 1.5 -- \qty{3}{K} from \qty{10}{km} to \qty{25}{km}

\subsection{Pressure Processing}
On board the GSB, the pressure readings from the four sensors are averaged, and outlier readings are detected and removed from the average. At least two sensors must agree to report a valid pressure, although in practice all four readings are used the vast majority of the time. The pressure sensor is unique in that it reports a single value per communication (typically every 1 to 7 minutes), rather than one every \qty{10}{s}. This exploits the fact that GNSS altitude is reported every \qty{10}{s}, and the difference between barometric altitude and GNSS altitude varies little on the timescale of a single transmission, which is several minutes. This difference is smoothed with a Savitzky–Golay filter over a 14 minute window and then interpolated to the altitude timestamps, which are used to calculate pressure values at those times. This approach allows us to minimize overall data downlink size while still reporting accurate pressure values, and is responsible for the smooth nature of the differences seen the pressure and altitude plots in Fig. \ref{fig:ucsd_flights}. 

\subsection{Comparison to WMO Requirements}
Our data resolution and calibration uncertainty requirements were designed to exceed the performance specifications in WMO OSCAR 2.1 Global Numerical Weather Prediction and Real-time Monitoring \citep{WMO_OSCAR_GlobalNWP}. The full list includes goal, breakthrough, and threshold values within different atmospheric layers, but the most stringent (goal) requirements across all layers are: wind vector -- \qty{0.5}{m/s}; pressure -- \qty{0.5}{hPa}; temperature -- \qty{0.5}{K}; and relative humidity -- 2\%\,RH. 

The relative humidity requirements apply only to the troposphere and are taken from Table 12.B.3 of the WMO Guide to Instruments and Methods of Observations Volume I – Measurement of Meteorological Variables \citep{WMO2025_CIMO}, rather than OSCAR 2.1. This is for the following reason, as stated in the WMO Guide: ``The OSCAR/Requirement...refer(s) to specific humidity, but this leads to far too stringent limits on uncertainty in layers where relative humidity is very low in the lower and middle troposphere. So values are shown as approximately equivalent relative humidity(.)"

These numbers may be compared, with important caveats, to the values in Table \ref{tab:sensor_specs}. The WMO uncertainty numbers are defined as the observation root mean square error (RMSE) at the $1\sigma$ level, and so direct comparison requires dividing the calibration uncertainty numbers by a factor of two. Furthermore, the sounding reproducibility numbers provide an overestimate of the true observation differences due to the separation between the two GSBs during flight. While noting that none of the numbers in Table \ref{tab:sensor_specs} provides a direct assessment of the true observation RMSE value, we find that our internal uncertainty characterization is consistent with meeting or exceeding the most stringent WMO requirements. 

\subsection{Quality Control Checks}\label{ssec:quality_control_checks}
All WindBorne GSB data is quality control (QC) checked before the data is used in different components within the company or shipped out to customers. All reported variables (altitude, pressure, wind components, temperature, and relative humidity) have different levels of QC applied to them. The QC checks are applied in real time, not post-processed, to allow the highest quality of data to be readily available.

One of the most important QC checks is applied to the initial ascent data to determine if our sensor package has deployed correctly. Unsuccessful deploy of the sensor package was a more common issue in the early times of developing the GSB platform. Now, this is less of a problem and typically occurs for fewer than 5\% of flights per month. If the sensor package has not deployed correctly, this can lead to erroneous temperature and humidity data. The QC algorithm for this check is fully automated; however, there are fringe cases that need to be reviewed by a human from time to time. The algorithm examines the temperature data during the first identified initial ascent to determine if the sensor package deployed correctly. If deployment of the sensor package failed, all temperature and relative humidity data are flagged as bad. However, wind data are still valid and collected.

One of the QCs for relative humidity is a comparison between a temperature sensor attached to the humidity sensors (humidity sensor temperature) and the ambient temperature measured from the thermistor. Since our humidity sensors are shielded in a package to protect them from environmental conditions, they can be impacted by solar radiation heating up the package. To account for the impacts from solar radiation, the ambient temperature from the thermistor is used to correct the relative humidity measurement, as described in Sec. \ref{sec:data_processing}\ref{ssec:humidity_processing} and Eq. \ref{eq:corrected_rh}. However, this correction can become large when the differences between humidity sensor temperature and ambient become large. To ensure reliability of relative humidity observations, QC flags data as invalid when the difference between humidity sensor temperature and ambient temperature exceeds acceptable thresholds. The thresholds are dependent on different conditions like ascent rate and solar elevation.

\begin{figure}
 \noindent\includegraphics[width=0.9\textwidth]{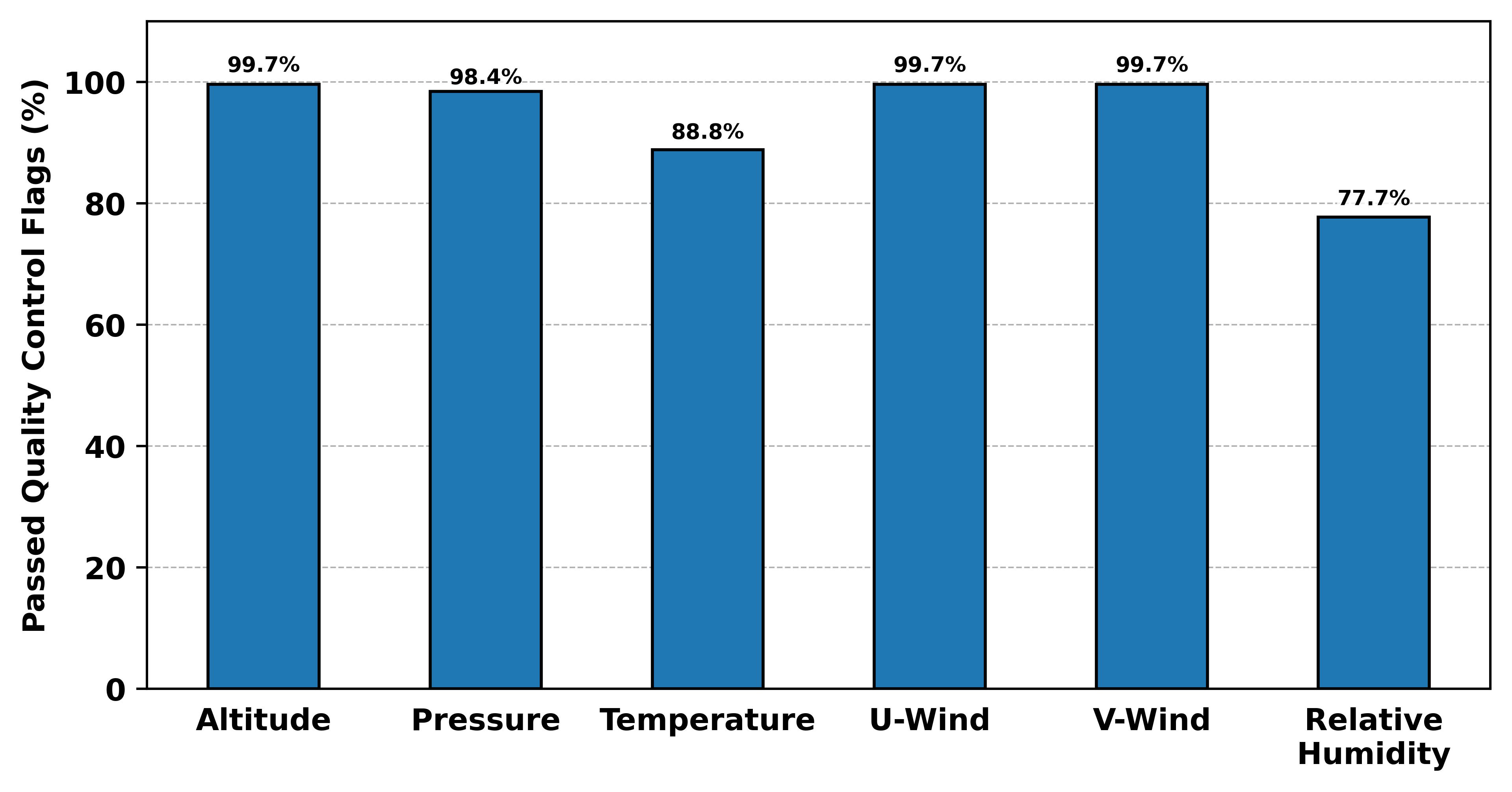}\\
 \caption{Observation validity percentage for all profiling data spanning 2024/7/1 through 2025/6/30.}\label{fig:GSB_QC}
\end{figure}

An example showing the percentage of GSB measurements that pass each variable's quality control is provided spanning 1 July 2024 through 1 July 2025 (Fig. \ref{fig:GSB_QC}). Since altitude, wind speed \textit{u}, and wind speed \textit{v} are measured using GPS location changes, their sensor QC checks are the same. These variables pass the QC check at a high percentage ($\approx 99 \%$), and the few data points flagged are usually due to GNSS jamming and location spoofing. Pressure measurements are just below the previous observations with an acceptance percentage of $\approx 98 \%$. Temperature and humidity measurements are lower in acceptance percentage. For temperature, the largest source of data loss is initial deployment of the sensor package failing. Humidity has the lowest acceptance percentage and there are two primary causes, both related to the temperature correction in Eq. \ref{eq:corrected_rh}. First, humidity must be invalid whenever temperature is invalid. Second, a relatively large number of humidity measurements are flagged due to the large difference between humidity sensor temperature and the ambient temperature, as discussed in the previous paragraph. This large temperature difference occurs during daytime due to solar impacts and when the balloon remains at a constant altitude with low sensor aspiration. During profiling, the sensor package has better aspiration leading to smaller temperature differences. Both of these sources of invalid data are expected to occur less frequently in the future, as the rate of failed sensor package deploys has decreased, and a new shield that leads to significantly less humidity sensor solar heating is expected to become standard in Q1 of 2026.

\section{Sounding Reproducibility}

\begin{figure}[t]
 \noindent\includegraphics[width=\textwidth]{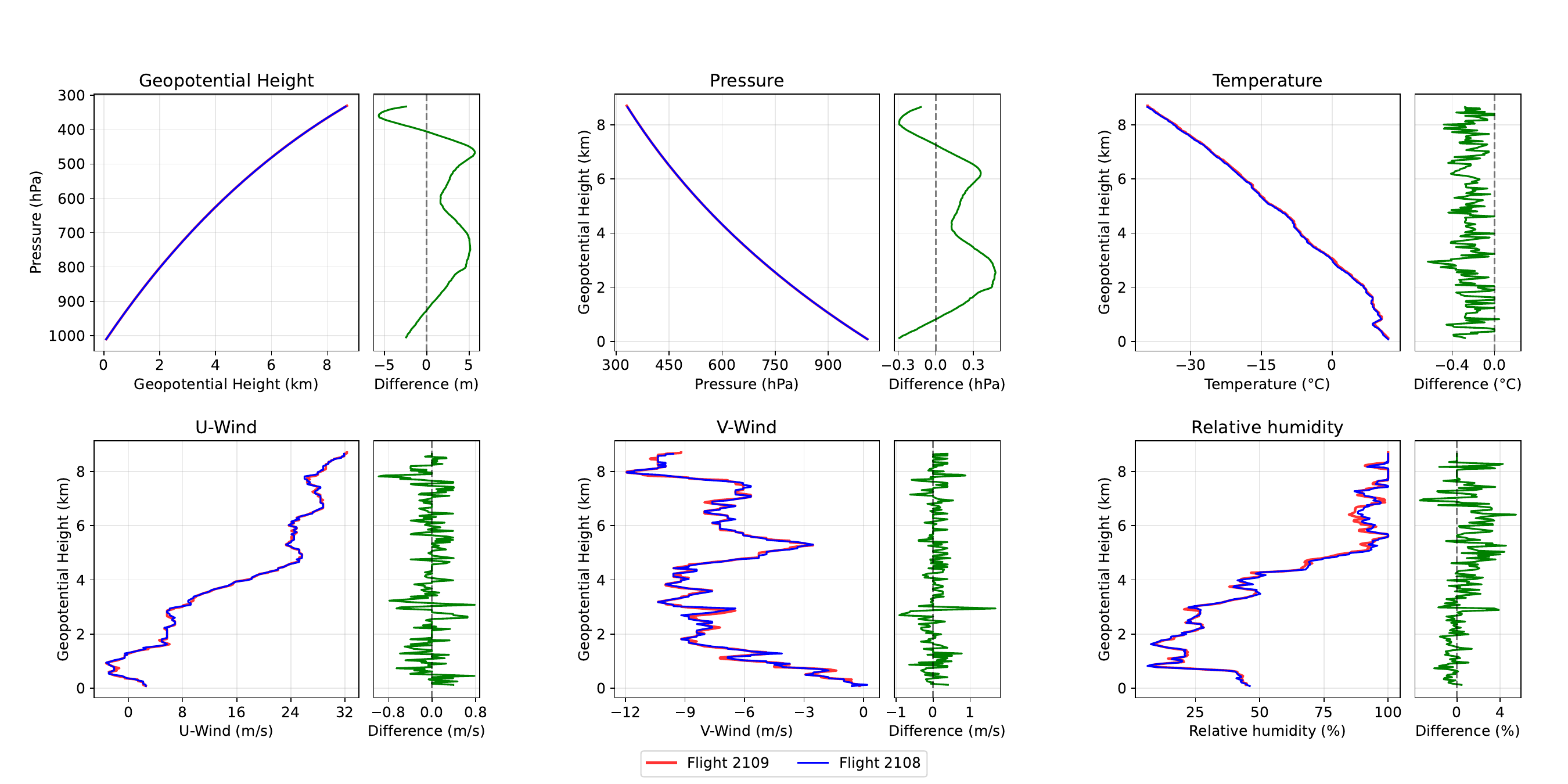}\\
 \caption{Comparison of initial ascent profiles from two simultaneously launched GSB-60s. All variables are matched by interpolation at equivalent geopotential height, except for geopotential height, which is matched at equivalent pressure. The initial ceiling for both flights was \qty{9}{km}.}\label{fig:simul}
\end{figure}

We have assessed sounding reproducibility by simultaneously launching two balloons side-by-side and comparing the data during their first ascent profiles. Data from one such comparison is shown in Figure ~\ref{fig:simul}. The reproducibility statistics found in Table \ref{tab:sensor_specs} are generated by combining data from five side-by-side launches, two with GSB-60 envelopes, which have an initial ceiling of about \qty{10}{km}, and three with GSB-80 envelopes, which have an initial ceiling of about \qty{19}{km}. During the ascent, the balloons drift apart both vertically and horizontally by up to several kilometers. We do not generally observe a significant increase in noise in the matched sensor data as a function of increasing separation with the exception of wind speed and relative humidity, both of which are noisier by about 20\% when comparing flights with less separation to those with higher separation. The root mean square difference between flights for each sensor variable, aggregated across all five flights, are presented in Table \ref{tab:sensor_specs}.

\section{Radiosonde Comparison}\label{sec:radiosonde_comparison}

\begin{figure}[t]
 \noindent\includegraphics[width=\textwidth]{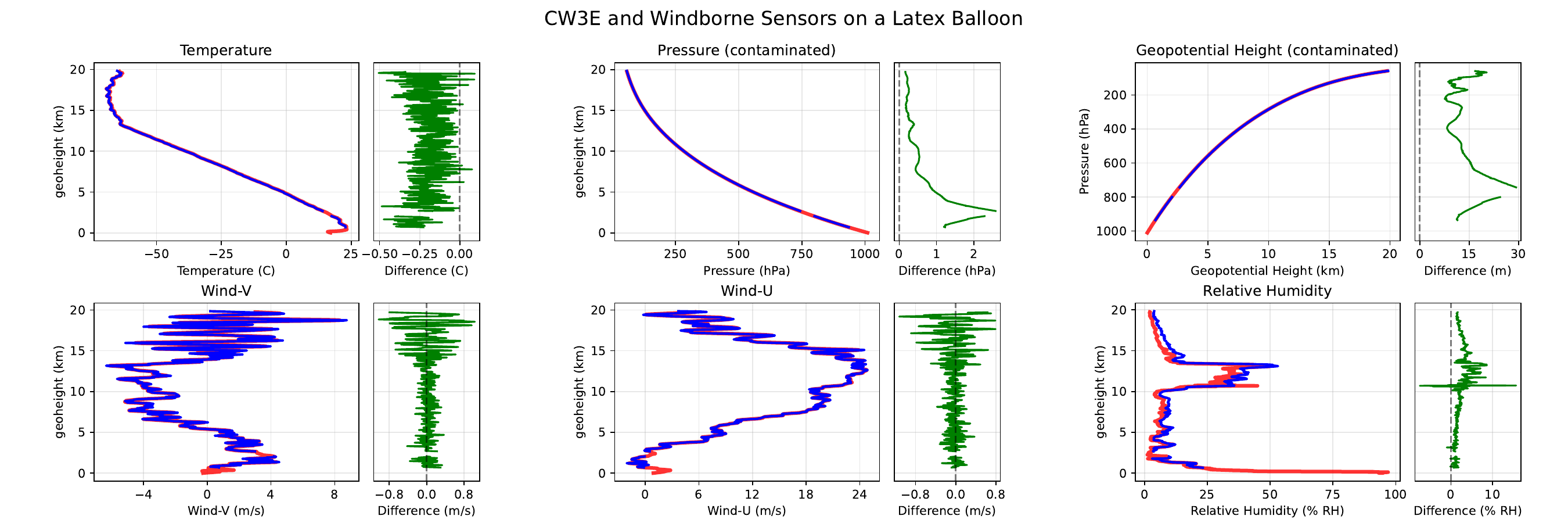}\\
 \noindent\includegraphics[width=\textwidth]{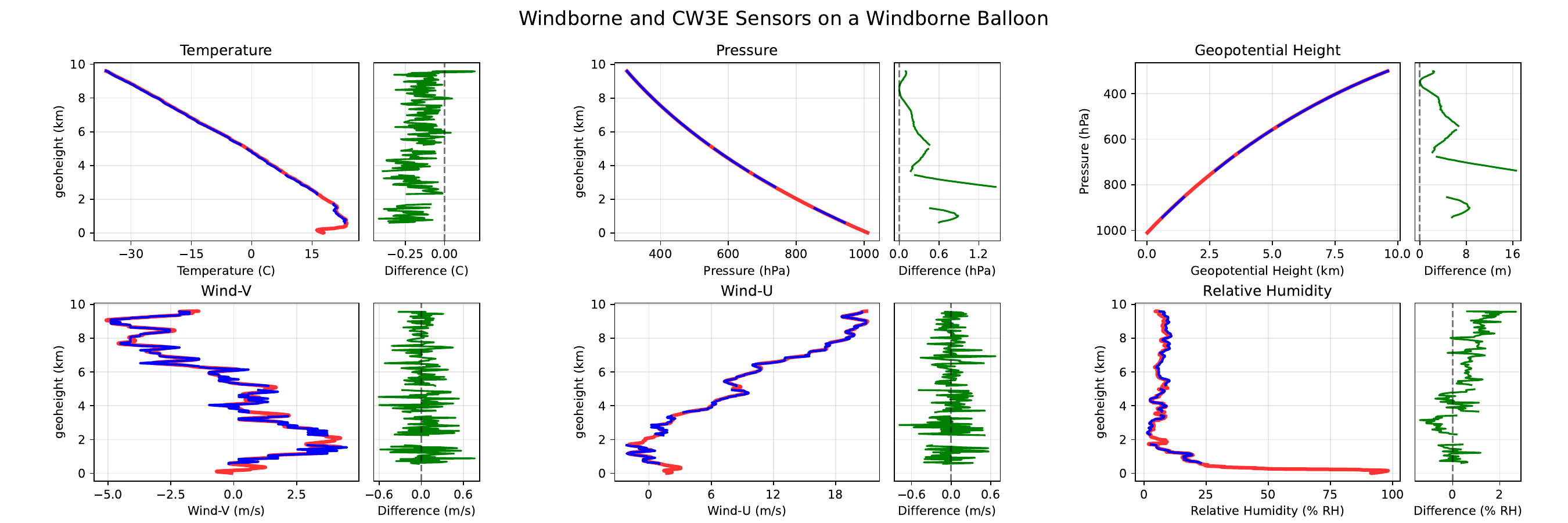}\\
  \noindent\includegraphics[width=\textwidth]{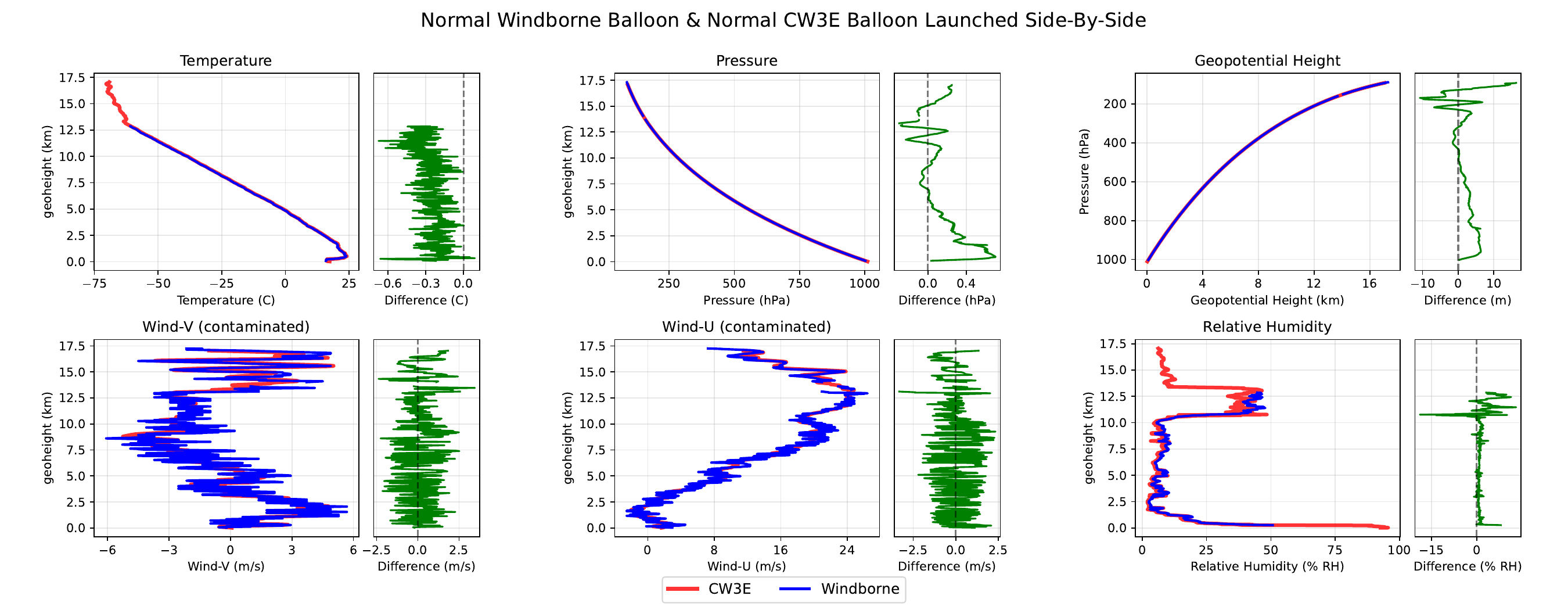}\\
 \caption{Comparison of initial ascent profiles from a WindBorne GSB-60 and a latex balloon with a Vaisala RS41-SGP radiosonde, simultaneously launched from the Scripps Pier in San Diego, CA. All variables are matched by interpolation to equivalent geopotential height, except for geopotential height, which is matched by pressure. Plots labelled `contaminated' have known sources of error, described in the text.}\label{fig:ucsd_flights}
\end{figure}

In May 2025 WindBorne performed an intercomparison study with radiosondes in partnership with the Center for Western Weather Extremes\footnote{https://cw3e.ucsd.edu/}, a part of the Scripps Institution of Oceanography at University of California San Diego. Soundings were taken with WindBorne sensors and Vaisala\footnote{https://www.vaisala.com/} RS41-SGP radiosondes, and were compared in three separate configurations:
\begin{enumerate}[label=Flight \arabic*., leftmargin=2cm]
    \item A latex balloon with WindBorne sensors and a radiosonde attached.
    \item A WindBorne balloon with WindBorne sensors and a radiosonde attached.
    \item A WindBorne balloon with WindBorne sensors and a latex balloon with a radiosonde launched side-by-side.
\end{enumerate}
The data from the first ascent profile for each flight is shown in Fig. \ref{fig:ucsd_flights}. All sensor packages on all flights successfully collected data. Only data that passed quality control are shown in the Figures. In matching wind speed data by altitude, we have accounted for the difference in height between the measuring device and the balloon, as the radiosonde is on a \qty{55}{m} line and the WindBorne GNSS receiver is right below the balloon. Flight 1 is an exception: the WindBorne unit was attached on a string about \qty{2}{m} below the radiosonde. For this flight we applied a low-pass filter to the WindBorne wind speed data, with approximately the same frequency cutoff as the radiosonde data, to remove the pendulum effect.

The comparison study necessitated altered configurations for some flights, which modified the sensor readings and resulted in two known sources of contamination. 1) The WindBorne wind speed readings in Flight 3 have elevated noise. This is because the balloon was given less initial ballast to allow it to achieve a higher altitude during its first ascent profile, which also caused the envelope to turn sideways and the GNSS receiver to sway. 2) The WindBorne pressure readings on Flight 1 are biased high, This is likely due to poor ventilation in the packaging, which was altered to attach it to the radiosonde. 

The differences shown in the uncontaminated comparisons generally show good agreement with the performance specifications presented in Table \ref{tab:sensor_specs} and the performance specifications for the RS41-SGP radiosonde\footnote{https://docs.vaisala.com/v/u/B211444EN-J/en-US}. The largest humidity differences occur at an altitude of \qty{10.7}{km} in flights 1 and 3. Here the radiosonde data shows a large oscillation in relative humidity over approximately \qty{90}{m}, likely corresponding to the radiosonde passing rapidly through thin clouds, whereas the WindBorne data is significantly smoother. At this altitude \qty{90}{m} corresponds to 24–-\qty{39}{s} of WindBorne data, which is just a few 10-second sampling periods and only 0.8–-1.2 time constants, so the humidity sensor does not fully resolve these oscillations. These high ascent rates occur mainly during the GSB's first ascent and are much faster than a typical sounding ($\approx$\qty{1}{m/s}), which would better capture such fine-scale structure. Smaller but notable differences may also be found in the temperature and humidity data: the temperature readings are consistently different by a couple tenths of one degree, and the humidity data also shows a small but persistent difference, which is discussed further at the end of Sec. \ref{sec:ERA5}. 

\section{Long-term ERA5 Comparison} \label{sec:ERA5}
A year-long evaluation has been completed comparing WindBorne GSB data to ECMWF Reanalysis v5 (ERA5; \cite{hersbach2020era5}). ERA5 was chosen since it provides continuous data over our year-long evaluation time period and the modeling system remains frozen, meaning no change in potential ERA5 biases as a result of version changes to either the NWP model or data assimilation system. Hourly ERA5 data are interpolated to the WindBorne GSB locations. Bilinear interpolation (horizontal location), log-space pressure interpolation (vertical location), and linear time interpolation are used to obtain ERA5 data at a specific point in space and time. The evaluation time period spans from 1 July 2024 through 30 June 2025 and the spatial locations are shown in Fig. \ref{fig:spatialCov_Varcounts}.

\begin{figure}[t]
 \noindent\includegraphics[width=\textwidth]{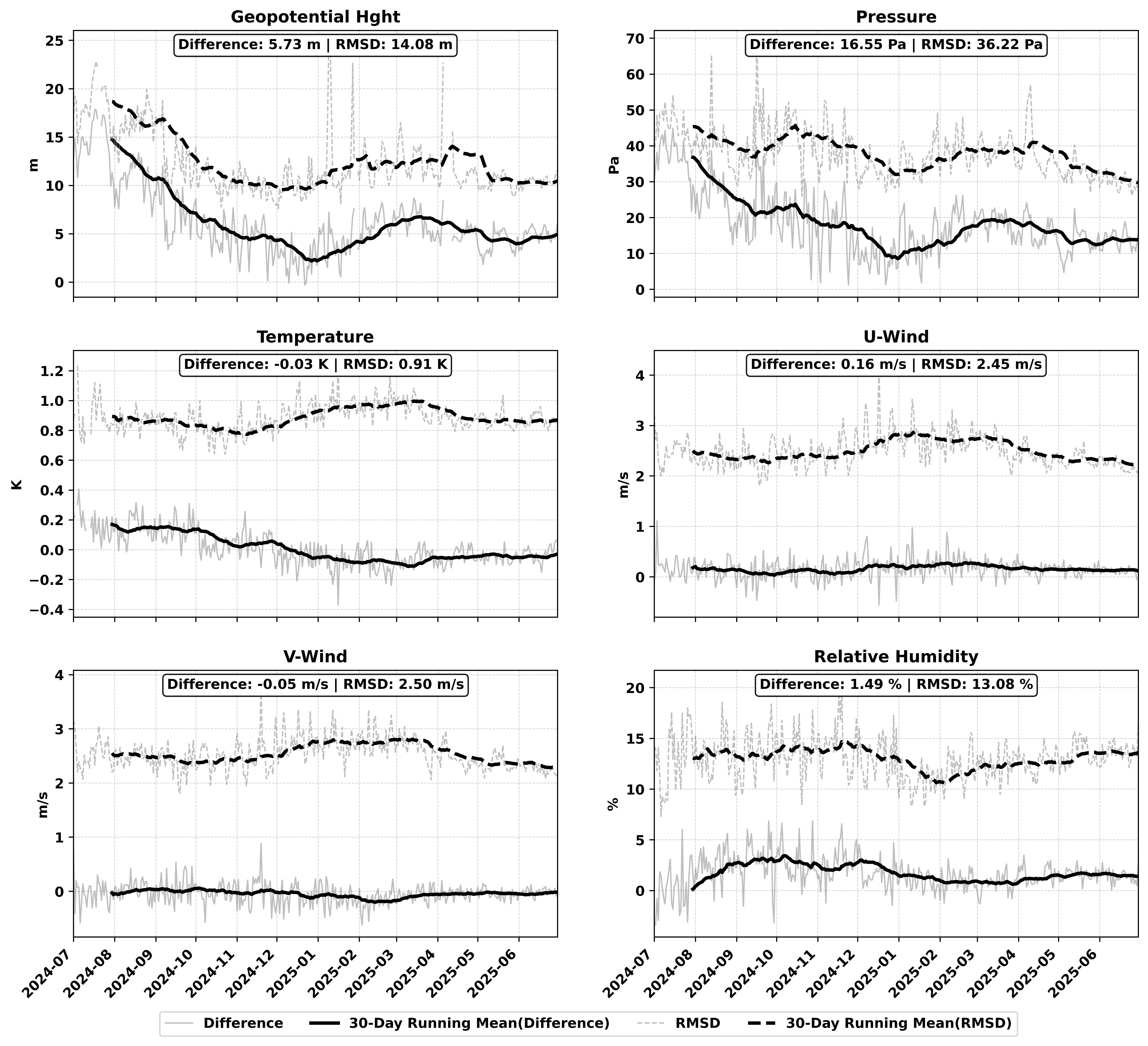}\\
 \caption{Daily differences between WindBorne to ERA5. Data is binned daily for computing the mean differences and root-mean-squared differences (RMSDs). A 30-day running mean is applied to both the differences and the RMSD to reduce the noise and smooth the signal.}\label{fig:timeseries_vs_era5}
\end{figure}

Daily mean differences and root-mean-squared differences (RMSDs) are computed for geopotential height, pressure, temperature, wind components, and relative humidity over the year-long evaluation period (Fig. \ref{fig:timeseries_vs_era5}). Both geopotential height and pressure have similar mean difference patterns and both have a positive differences of $\approx$\qty{5}{m} and $\approx$\qty{16}{Pa} when compared to ERA5. Since the GSBs obtain their altitude (convert to geopotential height) via GNSS, we are likely seeing the positive pressure difference projected onto the geopotential height since pressure is used in the height interpolation. For temperature, there is a positive difference of $\approx$\qty{0.1}{K} when compared to ERA5 that switches to a negative difference of $\approx$\qty{0.1}{K} after December 2024. This switch in sign is possibly due to the large expansion in launch site and associated data coverage that occurred in late 2024 and early 2025. There is a slight positive difference for \textit{u} (zonal) wind while \textit{v} (meridional) wind differences are more centered around zero. Both wind components have similar RMSD values computed over the year-long time period. There is a slight increase in the \textit{u} wind difference and RMSD compared to ERA5 spanning December 2024 thru April 2025. Due to our GSB data mainly being located in the Northern Hemisphere, this could be linked to sampling the strong polar jet stream that is usually observed during the wintertime months. Lastly, the relative humidity stats show a slight positive difference over the entire year-long evaluation with RMSD staying consistently around 12\% value. 

\begin{figure}[t]
 \noindent\includegraphics[width=\textwidth]{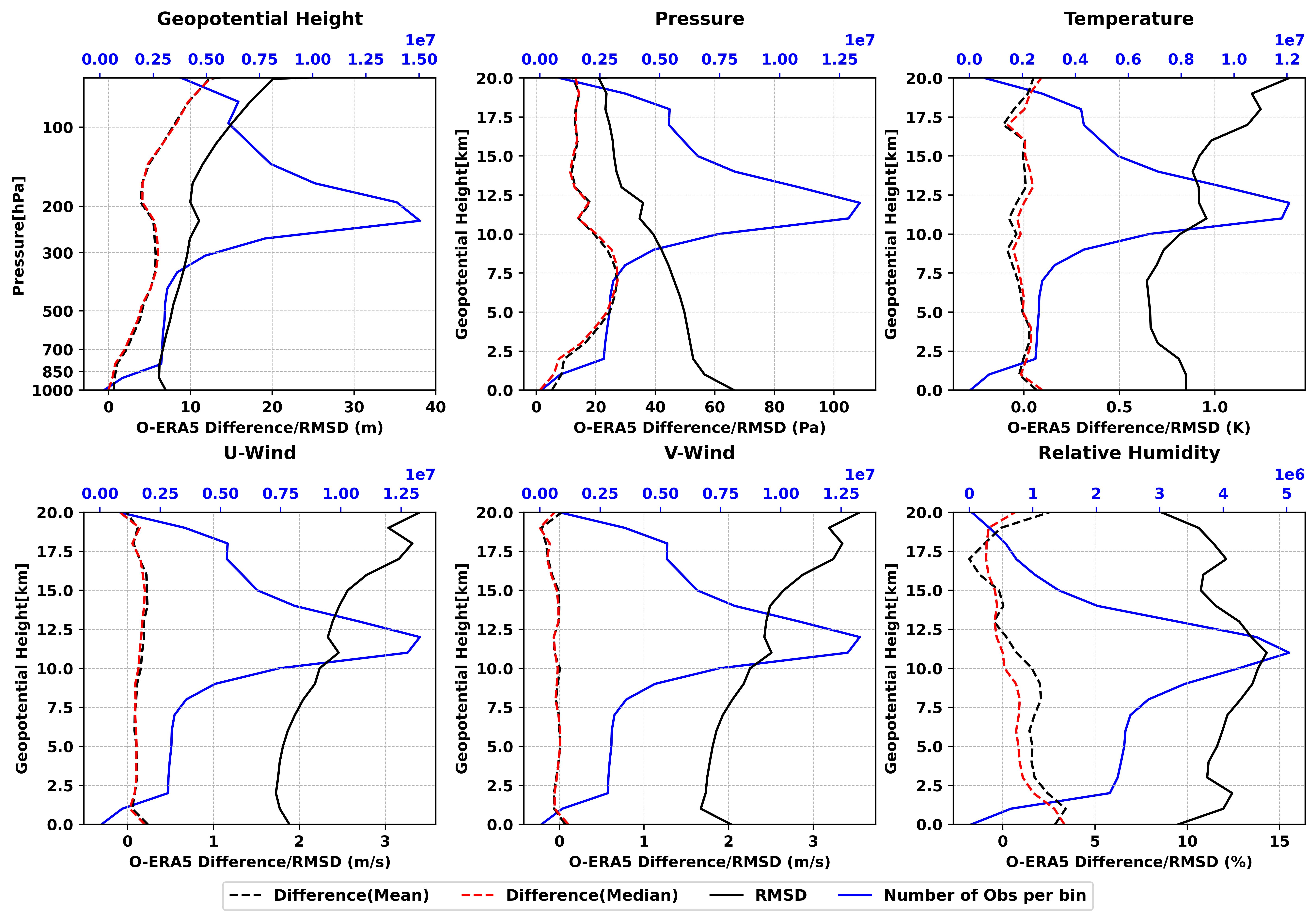}\\
 \caption{Vertical comparison of WindBorne to ERA5. Data is binned by geopoential height for all variables except for geopoential height stats. Mean (dashed black line) and median (dashed red line) differences is computed over these vertical bins along with RMSD (solid black line). Additionally, the number of samples in each bin are shown (blue line).}\label{fig:era5_vs_pressure}
\end{figure}

To better understand how the daily statistics are distributed in the vertical, departures between ERA5 and the GSB data are binned in the vertical height coordinate (Fig.\ref{fig:era5_vs_pressure}). Geopotential height is used for binning all GSB variables except geopotential height, which uses pressure as the vertical binning coordinate. Both geopotential height and pressure have a positive bias extending up for all the vertical bins. For temperature, in the altitudes with the most observations (\qty{7.5}{km} -- \qty{12.5}{km}) there is a small negative difference, but outside that region the differences compared to ERA5 are near zero. There is a positive difference that is largest at \qty{14}{km} in the \textit{u} wind extending throughout the vertical profile, however v wind has a near zero difference. Relative humidity has a positive mean difference below \qty{12}{km} but above this level the differences are near zero or slightly negative. Since our balloons profile from $\approx$\qty{12}{km} down to the surface, this positive bias in relative humidity is within our profiling dataset and not our cruising dataset. The RMSD for temperature increases around \qty{10}{km} or \qty{200}{hPa}. Additionally, the RMSD for the wind components is also increasing, more gradually, up above \qty{10}{km}.  The increase in RMSD for these fields starting at around \qty{200}{hPa} might be linked to sampling of data points near the tropopause or in the lower stratosphere, a region global models have been shown to struggle in representing \citep{shepherd2018report,bland2021characterising,kruger2022vertical,laloyaux2022deep}. The overall statistical patterns and magnitudes show similarities to studies comparing radiosondes observations with ERA5 \citep{huang2021evaluation,kuchinskaia2022era5,hassan2023assessment}.

\begin{figure}[t]
 \noindent\includegraphics[width=\textwidth]{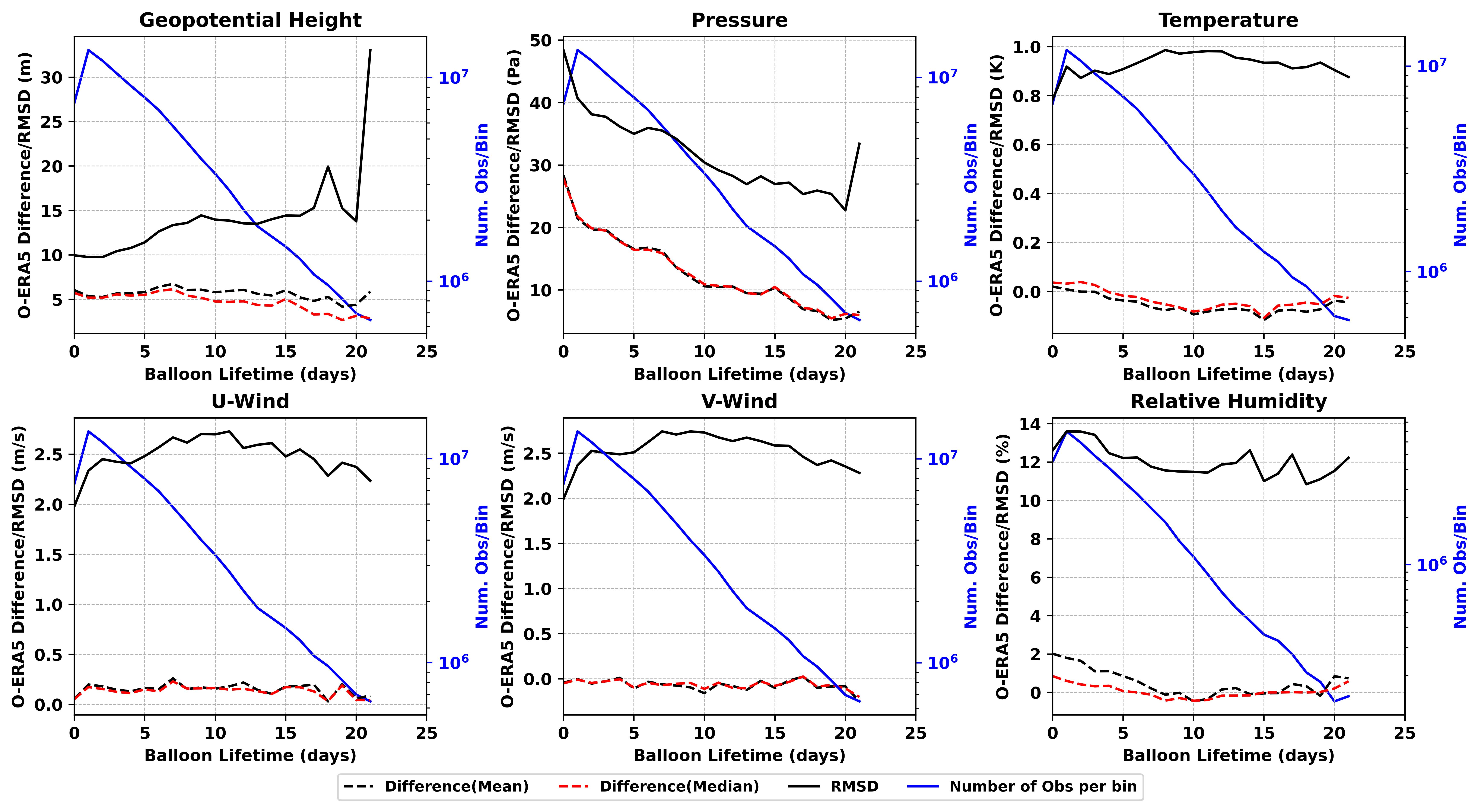}\\
 \caption{Balloon lifetime comparison of WindBorne to ERA5. Data is binned by balloon flight time since launch. Mean (dashed black line) and median (dashed red line) bias is computed over these lifetime bins along with RMSE (solid black line). Additionally, the number of samples in each bin are shown (blue line).}\label{fig:era5_vs_lifetime}
\end{figure}

Since GSBs are balloon platforms that last more than one ascent, it is important to determine the sensor quality throughout their lifetimes. This is achieved by binning GSB and ERA5 departures based on the balloon flight time since launch (Fig. \ref{fig:era5_vs_lifetime}). The average lifetime of a GSB is $\approx$7 days over this year-long evaluation period; however, with GSB modifications this average has increased over the course of this evaluation period. Geopotential height has a constant difference compared to ERA5 of around \qty{5}{m} while pressure has a decreasing difference with balloon lifetime. This is likely linked to GSBs spending more time at constant high altitudes (cruising altitudes) later in flight where differences in pressure are smaller compared to near the surface over the same height difference. Temperature is mostly stable compared to ERA5, with the exception of a slight negative temperature difference that develops with increasing lifetime. A slight positive difference is observed over the GSB lifetime in the u wind, agreeing with the vertical profile statistics (Fig. \ref{fig:era5_vs_pressure}), while v-wind is near-zero. Lastly, the differences observed for relative humidity decreases due to less profiling at longer lifetimes. Overall, it appears that the sensors are not degrading over their lifetimes while flying on average up to a week in the atmosphere. Over the lifetime of GSBs, profiling becomes less of a flight strategy and cruising at constant altitude (commonly around \qty{12}{km}) becomes more common. This leads later in-flight stats to compare more closely with the statistics found in the vertical plots (Fig. \ref{fig:era5_vs_pressure}) at higher levels.

\begin{figure}[t]
 \noindent\includegraphics[width=\textwidth]{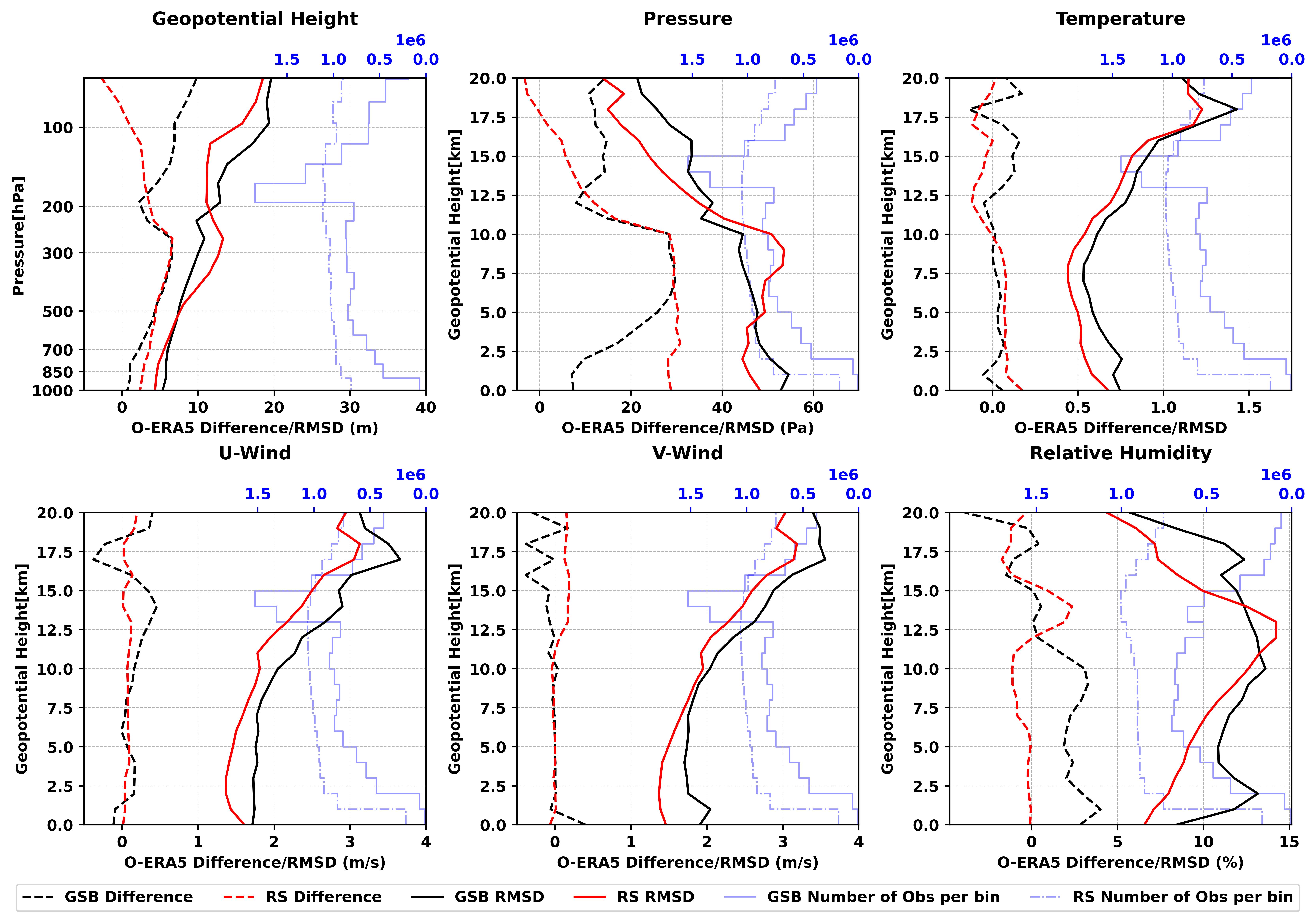}\\
 \caption{Vertical comparison of WindBorne (GSB; black) and radiosondes (RS; red) to ERA5 in November 2025. Data is binned by geopoential height for all variables except for geopoential height stats. Mean differences (dashed lines) are computed over these vertical bins along with RMSD (solid lines). Additionally, the number of samples in each bin are shown for WindBorne (blue solid line) and radiosonde (blue dashed).}\label{fig:GSB_vs_radiosonde}
\end{figure}

Since the end of June 2025, the last date in our year-long evaluation period, there have been several upgrades to data processing and hardware, most notably an update to the radiative temperature bias correction discussed in Section \ref{sec:radiative_bias}. To assess performance with these updates, we computed statistics comparing GSB observations to ERA5 over November 2025 (Fig. \ref{fig:GSB_vs_radiosonde}). To understand if the differences from ERA5 are unique to our platform, we also included comparisons between ERA5 and the National Centers for Environmental Information (NCEI) high-resolution US radiosonde dataset \citep{NWS2021}. This dataset provides high temporal and spatial resolution observation data points. It is important to note that radiosondes are currently assimilated into the ERA5 final analyses ( \cite{hersbach2020era5}), while GSB data are not, and thus radiosondes would be expected to show better agreement with ERA5. However, this provides the ability to see if both observing platforms have similar differences when compared to ERA5.

The mean differences for geopotential heights are similar between GSBs and radiosondes for geopotential height up to \qty{200}{hPa}, after which the lines diverge. For pressure, GSBs have lower differences in the lower troposphere compared to radiosondes, but the two lines converge and are similar up to \qty{12}{km}, where the two lines diverge. The mean difference values are similar between GSBs and radiosondes for temperature up to around \qty{12}{km}, where GSBs have a positive difference and radiosonde mean differences are different. For both wind components, the mean differences are similar between GSBs and radiosondes; however, GSBs have slightly larger positive differences for zonal (u) wind when compared to ERA5. Relative humidity has the largest discrepancies between GSBs and radiosondes. The mean differences for GSBs are shifted slightly positive throughout the troposphere compared to radiosondes. The RMSD lines are quite similar in the patterns between GSBs and radiosondes; however, the magnitudes are slightly shifted higher for GSBs. Again, this should be expected due to the assimilation of radiosonde data into ERA5 analyses.

There are similarities between ERA5 differences found when comparing between radiosondes and WindBorne GSBs. Most of the RMSD profiles follow a similar pattern. The slight shift to lower RMSD values for radiosondes is potentially linked to the fact that they are assimilated into ERA5 analyses. Additionally, both radiosondes and GSBs highlight the difficulties ERA5 has in representing the humidity gradient near the tropopause \citep{kohler2024relative}. There are a couple of areas where GSB mean differences have different patterns than radiosondes. The pressure in lower stratosphere is one area where the mean differences between the two observing platforms are slightly different. The source driving the mean differences is unknown at this time. The increased difference observed near \qty{13}{km} is likely due to GSB platform continuously sampling the upper-level jet stream when in cruising mode. Since the evaluation period was over November (Fig. \ref{fig:GSB_vs_radiosonde}), ERA5 could be under representing the strong polar jet stream that is enhanced during the winter. 

Finally, GSB relative humidity shows a positive mean difference throughout the troposphere, consistent with the direction of the differences observed in Sec. \ref{sec:radiosonde_comparison}. We have investigated potential instrumental causes, including the calibration procedure, incorrect adjustments for solar heating, contamination from sensor packaging materials, and biases from moisture accumulation, but have not found evidence that would explain the observed differences. A portion of the discrepancy relative to ERA5 may stem from sampling effects. In-situ readings exhibit significantly more extreme RH values than the reanalysis, which tends to be smoothed. The size of this effect exceeds the $\approx 2$\% -- 3\%\,RH average difference shown in Fig. \ref{fig:era5_vs_lifetime} by a factor of several and likely explains the divergence between the mean and median lines. This disparity must be carefully accounted for when interpreting these differences.

\begin{figure}
    \centering
    \includegraphics[width=1\linewidth]{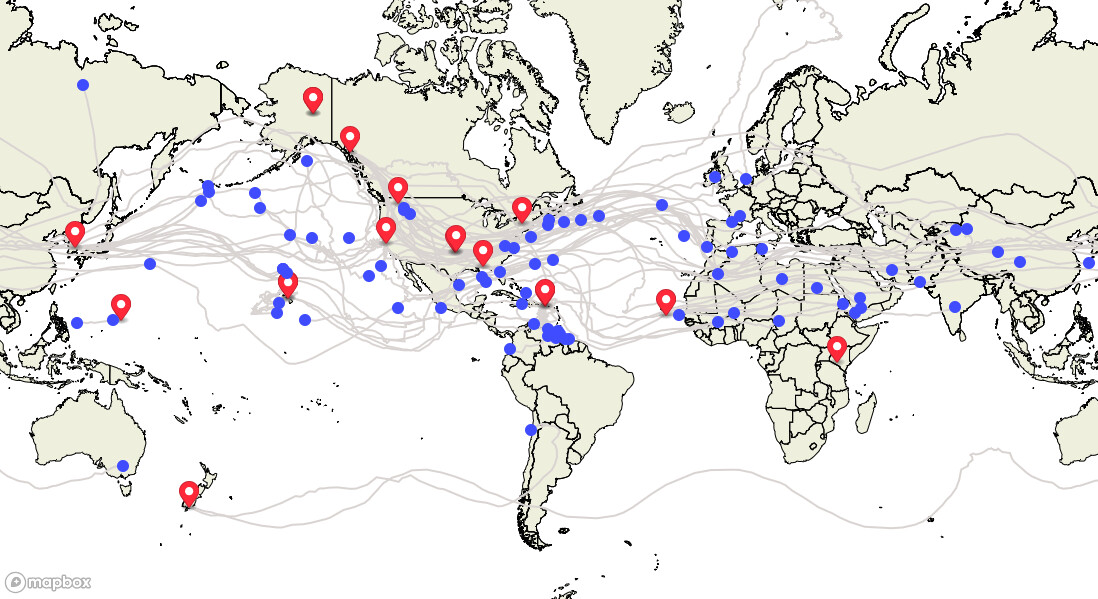}
    \caption{Flight paths (grey) and locations (blue dots) of all 81 GSBs that were aloft as of 00 UTC 21 January 2026. Launch locations are shown as red markers.}
    \label{fig:flight_paths}
\end{figure}
\section{Conclusions}


WindBorne System's long-duration controllable balloon-based observation platform, known as the Global Sounding Balloon, has been introduced in this manuscript. The performance of atmospheric observations collected by GSBs have been characterized by comparing to adjacently flown, industry-standard radiosonde systems and by comparing to global re-analyses of the atmospheric state over a year-long period.  Observations collected by GSBs have been shown to provide similar, and in some cases, superior performance to industry-standard radiosonde observations. WindBorne Systems currently flies a growing constellation of GSBs that is continuously collecting atmospheric observations throughout the troposphere over large segments of the world. A snapshot of the constellation as of 00 UTC 21 January 2026 is shown in Fig. \ref{fig:flight_paths}. While observations presented in this study are primarily over the northern hemisphere, as the constellation expands, observations will be collected throughout the world.  By the end of 2026, WindBorne Systems expects the constellation size to quadruple to an average of roughly 400 GSBs aloft at any given time. The meteorological sensor suite will also undergo upgrades in 2026, including an improved shielding design that will significantly increase the fraction of valid humidity data.

WindBorne Systems has participated in multiple field projects during which hundreds of GSBs were flown to test, advance and demonstrate the capabilities of the GSB.  During August of 2021 and August of 2022, WindBorne Systems launched GSBs from Alaska and Svalbard to observe conditions in the Arctic region and support observation collecting during the THINICE field campaign \citep{thinice}. During the 2022, 2024, and 2025 Atlantic hurricane seasons, WindBorne Systems collected observations throughout the tropical Atlantic in coordination with NOAA Hurricane Research Division field campaign \citep{WindBorne-GFS}.  Additionally, during the 2022-23, 2023-24 and 2024-25 winter seasons, WindBorne Systems participated in the Atmospheric Rivers Reconnaissance field program by collecting observations throughout the northern Pacific Ocean \citep{ARRecon}.  

Data from many of the field campaigns was evaluated by NOAA's Environmental Modeling Center (EMC) to assess forecast impact by running the Global Forecast System (GFS) retrospectively with and without WindBorne observations.  WindBorne observations were shown to yield positive forecast impacts \citep{WindBorne-GFS}. During August of 2025, observations collected by WindBorne's Atlas constellation began being assimilated into operational forecast systems, including NOAA's Global Forecast System (GFS) and WindBorne's WeatherMesh forecast system. Further, sounding profiles are delivered to NOAA's National Weather Service (NWS) AWIPS system and are available on skew-T, log-P diagrams to weather forecasters to aid in understanding situational awareness. 

WindBorne observations and data have been used by the U.S. Air Force for meteorological purposes in  experimentation and operation, and are currently under evaluation for inclusion in the Global Air-Land Weather Exploitation Model (GALWEM). 

\clearpage
\acknowledgments
The authors thank UC San Diego and the Scripps Institution of Oceanography for the use of the Ellen Browning Scripps Memorial Pier during the radiosonde comparison balloon launches and Ethan Morris for his help in launching the balloons.  The authors are especially grateful to Xingren Wu at NOAA's Environmental Modeling Center, whose continued efforts in performing forecast impact evaluations led to the operational use of the observations presented here within the GFS.  The authors acknowledge Bruce Ingleby and Magnus Lindskog at the European Centre for Medium-Range Weather Forecasting and Matthew Fry at the United Kingdom Meteorological Office for their invaluable comparisons of GSB observations with the atmospheric analyses that led to discoveries and ultimately resolutions of GSB data processing errors and an understanding of impacts on forecasts.  The authors also thank Jay Martinelli, Samuel Childs, and the Air Force 16th Weather Squadron for their time and evaluation of WindBorne's observational data, as well as Lt Col Andrew Travis and Lt Col Joseph Nash for making such evaluations possible.

Research and development of the Global Sounding Balloon and sensor packages has been supported, in part, under funding from NOAA's Weather Program Office (grant NA21OAR4590396) and the Air Force Research Laboratory (contract FA86492099123).  Operations are supported, in part, by NOAA's National Mesonet Program and the Air Force Life Cycle Management Center.

%
%
\datastatement
All CW3E radiosonde data created or used during this study are openly available at \url{https://cw3e-datashare.ucsd.edu/CW3E_Radiosondes/USSIO/}. Due to confidentiality agreements, other supporting data can only be made available to bona fide researchers subject to a non-disclosure agreement. Details of the data and how to request access are available from WindBorne Systems at data@windbornesystems.com.

\clearpage
\bibliographystyle{ametsocV6}
\bibliography{references}

\end{document}